\documentstyle[12pt,epsfig,graphics]{article}
\begin{document}\bibliographystyle{plain}\begin{titlepage}
\renewcommand{\thefootnote}{\fnsymbol{footnote}}\hfill\begin{tabular}{l}
HEPHY-PUB 736/01\\UWThPh-2000-52\\CUQM-84\\hep-th/0101223\\January
2001\end{tabular}\\[.5cm]\Large\begin{center}{\bf ENERGY BOUNDS FOR THE
SPINLESS SALPETER EQUATION}\\\vspace{0.8cm}\large{\bf Richard L.
HALL\footnote[3]{\normalsize\ {\em E-mail address\/}:
rhall@mathstat.concordia.ca}}\\[.3cm]\normalsize Department of Mathematics
and Statistics, Concordia University,\\1455 de Maisonneuve Boulevard West,
Montr\'eal, Qu\'ebec, Canada H3G 1M8\\[0.7cm]\large{\bf Wolfgang
LUCHA\footnote[1]{\normalsize\ {\em E-mail address\/}:
wolfgang.lucha@oeaw.ac.at}}\\[.3cm]\normalsize Institut f\"ur
Hochenergiephysik,\\\"Osterreichische Akademie der
Wissenschaften,\\Nikolsdorfergasse 18, A-1050 Wien,
Austria\\[0.7cm]\large{\bf Franz F.~SCH\"OBERL\footnote[2]{\normalsize\ {\em
E-mail address\/}: franz.schoeberl@univie.ac.at}}\\[.3cm]\normalsize Institut
f\"ur Theoretische Physik, Universit\"at Wien,\\Boltzmanngasse 5, A-1090
Wien, Austria\vfill {\normalsize\bf Abstract}\end{center}\normalsize We study
the spectrum of the Salpeter Hamiltonian $H=\beta\sqrt{m^2+{\bf p}^2}+V(r)$,
where $V(r)$ is an attractive central potential in three dimensions. If
$V(r)$ is a convex transformation of the Coulomb potential $-1/r$ and a
concave transformation of the harmonic-oscillator potential $r^2$, then upper
and lower bounds on the discrete eigenvalues of $H$ can be constructed, which
may all be expressed in the form
$$E=\min_{r>0}\left[\beta\sqrt{m^2+\frac{P^2}{r^2}}+V(r)\right]$$for suitable
values of $P$ here provided. At the critical point $r=\hat r$ the relative
growth~to~the Coulomb potential $h(r)=-1/r$ must be bounded by ${\rm d}V/{\rm
d}h<2\beta/\pi.$\\[3ex]{\em PACS numbers\/}: 03.65.Ge, 03.65.Pm, 11.10.St
\renewcommand{\thefootnote}{\arabic{footnote}}\end{titlepage}

\normalsize

\section{Introduction}We study the (semirelativistic) ``spinless-Salpeter''
Hamiltonian\begin{equation}H=\beta\sqrt{m^2+{\bf p}^2}+V(r)\ ,\quad\beta>0\
,\label{Eq:SH}\end{equation} in which $V(r)$ is a central potential in three
spatial dimensions. The eigenvalue equation~of this operator is called the
``spinless Salpeter equation.'' This equation of motion arises as~a
well-defined standard approximation to the Bethe--Salpeter formalism
\cite{Salpeter51} for the description of bound states within a (relativistic)
quantum field theory and is arrived at by the following simplifying
steps:\begin{enumerate}\item Eliminate all timelike variables by assuming the
Bethe--Salpeter kernel that describes the interactions between the
bound-state constituents to be static, i.e., instantaneous; the result of
this reduction step is called the ``instantaneous Bethe--Salpeter equation''
or the ``Salpeter equation'' \cite{Salpeter52}.\item Neglect the spin of the
bound-state constituents, assume the Bethe--Salpeter kernel~to be of
convolution type (as is frequently the case), and consider merely
positive-energy solutions $\psi,$ in order to arrive at the so-called
``spinless Salpeter equation'' $H\psi=E\psi,$ with a Hamiltonian $H$ of the
form (\ref{Eq:SH}). (For two particles, this form of the Hamiltonian $H$
holds only for equal masses $m$ of the bound-state
constituents.)\end{enumerate}(For a more detailed account of the reduction of
the Bethe--Salpeter equation to the spinless Salpeter equation, consult,
e.g., the introductory sections of Refs.~\cite{Lucha98O,Lucha98D}.) This wave
equation describes the bound states of spin-zero particles (scalar bosons) as
well as the spin-averaged spectra of the bound states of fermions.

In this paper we consider potentials which are at the same time convex
transformations $V(r)=g(h(r))$ of the Coulomb potential $h(r)=-1/r$ and
concave transformations of the harmonic-oscillator potential $h(r)=r^2.$ The
reason for this is that spectral information is known for these two ``basis''
potentials $h(r)$. Thus the class of potentials is those $V(r)$ that have a
dual representation$$V(r)=g^{(1)}\left(-\frac{1}{r}\right)=g^{(2)}(r^{2})\
,$$in which $g^{(1)}$ is convex (${g^{(1)}}''>0$) and $g^{(2)}$ is concave
(${g^{(2)}}''<0$). An example of a potential in this class
is\begin{equation}V(r)=-\frac{c_1}{r}+c_2\ln r+c_3r+c_4r^2\
,\label{Eq:pot-class}\end{equation}where the coefficients $\{c_{i}\}$ are not
negative and are not all zero. Thus tangent lines to the transformation
function $g(h)$ are of the form $ah+b$ and are either Coulomb
potentials~lying below $V$, or harmonic-oscillator potentials lying above
$V.$ This geometrical idea is the basis for our approach to the spectral
problem posed by $H.$ We recall the application of this idea to the
(nonrelativistic) Schr\"odinger problem in Sec.~\ref{Sec:NRET}. The general
envelope formalism for the derivation of upper and lower bounds on the
eigenvalues of the semirelativistic Salpeter Hamiltonian $H$ of
Eq.~(\ref{Eq:SH}) is established in Sec.~\ref{Sec:RET}.

It is fundamental to our method that we first know something about the
spectrum~of~$H$ in those cases where $V(r)$ is one of the basis potentials,
i.e., the Coulomb and the~harmonic oscillator. These two spectra are
discussed in Sec.~\ref{Sec:CHOP} below. In Sec.~\ref{Sec:FP} we look at the
example of the Coulomb-plus-linear potential.

\section{The Coulomb and harmonic-oscillator potentials}\label{Sec:CHOP}
\subsection{Scaling behaviour}Since the two basis potentials are both pure
powers, it is helpful first to determine what~can be learnt about the
corresponding eigenvalues by the use of standard scaling arguments. By
employing a wave function $\phi(cr)$ depending on a scale variable $c>0,$ we
find the following scaling rule for the eigenvalues corresponding to
attractive pure power potentials~$v\,{\rm sgn}(q)r^q.$ The
Hamiltonian$$H=\beta\sqrt{m^2+{\bf p}^2}+v\,{\rm sgn}(q)r^q$$has the (energy)
eigenvalues $E(v,\beta,m),$ where$$E(v,\beta,m)=\beta
mE\!\left(\frac{v}{\beta m^{1+q}},1,1\right),\quad q\ge -1\ .$$The scaling
behaviour described by the above formula allows us to consider the
one-particle, unit-mass special case $m=\beta=1$ initially, that is to say,
to work w.l.o.g.\ with the operator$$H=\sqrt{1+{\bf p}^2}+v\,{\rm
sgn}(q)r^{q}\ .$$

\subsection{Coulomb potential}\label{Subsec:CP}In the case of the Coulomb
potential $V(r)=-v/r$ it is well known \cite{Herbst77} that the Hamiltonian
$H$ has a Friedrichs extension provided the coupling constant $v$ is not too
large. Specifically, it is necessary in this case that $v$ is smaller than a
critical value $v_{\rm c}$ of the coupling constant:$$v<v_{\rm
c}=\frac{2}{\pi}\ .$$With this restriction, a lower bound to the bottom of
the spectrum is provided by Herbst's formula\begin{equation}
E_0\ge\sqrt{1-(\sigma v)^{2}}\ ,\quad\sigma\equiv\frac{\pi}{2}\
.\label{Eq:Herbst-bound}\end{equation}By comparing the spinless Salpeter
problem to the corresponding Klein--Gordon equation, Martin and Roy
\cite{Martin89} have shown that if the coupling constant is further
restricted by $v<\frac{1}{2},$ then an improved lower bound is provided by
the expression\begin{equation}E_0\ge\sqrt{\frac{1+\sqrt{1-4v^{2}}}{2}}\
,\quad v<\frac{1}{2}\ .\label{Eq:MR-bound}\end{equation}It turns out that our
lower-bound theory has a simpler form when the Coulomb eigenvalue formula has
the form of Eq.~(\ref{Eq:Herbst-bound}) rather than that of
Eq.~(\ref{Eq:MR-bound}). For this reason, we have derived from
Eq.~(\ref{Eq:MR-bound}) a family of Coulomb bounds which, by rather
elementary methods, is found~to read\begin{equation}E_0\ge\sqrt{1-(\sigma
v)^2}\ ,\quad v\le\frac{\sqrt{\sigma^2 -1}}{\sigma^2}<\frac{1}{2}\
.\label{Eq:NLB}\end{equation}All these (lower) bounds are slightly weaker
than the Martin--Roy bound (\ref{Eq:MR-bound}) but above the Herbst bound
(\ref{Eq:Herbst-bound}). We note that these functions of the coupling
constant $v$ are all monotone and {\it concave\/}.

\subsection{Harmonic-oscillator potential}\label{Subsec:HOP}In the case of
the harmonic-oscillator potential, i.e., $V(r)=vr^2,$ much more is
known~\cite{Lucha99Q,Lucha99A}. In momentum-space representation the operator
${\bf p}$ becomes a $c$-variable and thus, from the spectral point of view,
the Hamiltonian $H=\sqrt{1+{\bf p}^2}+vr^2$ is equivalent to the
Schr\"odinger operator\begin{equation}H=-v\Delta+\sqrt{1+r^2}\
.\label{Eq:SHam-HO}\end{equation}Since the potential $V(r)$ increases without
bound, the spectrum of $H$ is entirely discrete~\cite{Reed78}. We denote its
eigenvalues by ${\cal E}_{n\ell}(v),$ where $n=1,2,3,\dots$ ``counts'' the
radial states in~each angular-momentum subspace labelled by
$\ell=0,1,2,\dots.$ Below we shall either approximate the eigenvalues ${\cal
E}(v)$ analytically or assume them to be known numerically. The eigenvalues
${\cal E}(v)$ of such Schr\"odinger operators are {\it concave\/} functions
of the coupling constant $v$ \cite{Thirring90,Hall83}.

\subsection{The spectral comparison theorem}\label{Subsec:SCT}For the class
of interaction potentials given by (\ref{Eq:pot-class}) with the coefficient
of the Coulombic~term not too large, that is, for all potentials which
satisfy the constraint $\lim_{r\to 0}r^2V'(r)<2\beta/\pi,$ the
semirelativistic Salpeter Hamiltonian $H$ is bounded below and is essentially
self-adjoint \cite{Herbst77}. Consequently, the discrete spectrum of $H$ is
characterized variationally \cite{Reed78} and it follows immediately from
this that, if we compare two such Hamiltonians $H$ having the potentials
$V^{(1)}(r)$ and $V^{(2)}(r),$ respectively, and we know that
$V^{(1)}(r)<V^{(2)}(r),$ then we may conclude that the corresponding discrete
eigenvalues $E_{n\ell}$ satisfy the inequalities
$E_{n\ell}^{(1)}<E_{n\ell}^{(2)}.$ We shall refer to this fundamental result
as the ``spectral comparison theorem.'' In the more~common case of
nonrelativistic dynamics, i.e., for a (nonrelativistic) kinetic term of the
form $\beta{\bf p}^2/2m$ in the Hamiltonian $H$, a constraint similar to the
above would hold for the coefficient of a possible additional (attractive)
$-1/r^2$ term in the potential $V(r).$

\section{General envelope theory of Schr\"odinger operators}\label{Sec:NRET}
In nonrelativistic envelope theory \cite{Hall83,Hall84,Hall93}, if the
potential $V$ is a smooth transformation $V(r)=g({\rm sgn}(q)r^q)$ of the
power-law potential ${\rm sgn}(q)r^q$, the eigenvalues of $H=-\Delta+V(r)$
are approximated by\begin{equation}E_{n\ell}\approx\min_{r>0}
\left[\frac{P_{n\ell}^2(q)}{r^2}+V(r)\right].\label{Eq:EVs}\end{equation}The
numbers $P_{n\ell}(q)$ can be derived from the eigenvalues of $-\Delta+{\rm
sgn}(q)r^q$ \cite{Hall93}. If $g$ is~convex ($g''>0$), Eq.~(\ref{Eq:EVs})
yields lower bounds; if $g$ is concave ($g''<0$), the~results are upper
bounds.

As an immediate application we consider the (nonrelativistic) Schr\"odinger
Hamiltonian (\ref{Eq:SHam-HO}) for the Salpeter harmonic-oscillator problem
(\ref{Eq:SH}). Here we have $H=-v\Delta+V(r),$ with
$V(r)=\beta\sqrt{m^2+r^2};$ hence, the potential is a convex transformation
of a linear potential~and a concave transformation of a harmonic-oscillator
potential. We conclude therefore from~(\ref{Eq:EVs}):\begin{equation}
\min_{r>0}\left[v\frac{P_{n\ell}^2(1)}{r^2}+\beta\sqrt{m^2+r^2}\right]\leq
{\cal E}_{n\ell}(v)\leq
\min_{r>0}\left[v\frac{P_{n\ell}^2(2)}{r^2}+\beta\sqrt{m^2+r^2}\right];
\label{Eq:SEVF-HO}\end{equation}the numbers $P_{n\ell}(1)$ are given in
Table~1 of Ref.~\cite{Hall00}, and $P_{n\ell}(2)=2n+\ell-\frac{1}{2}$.
Interestingly the upper and lower bounds (\ref{Eq:SEVF-HO}) are equivalent to
the corresponding bounds obtained~in~Ref.~\cite{Lucha99A}; however, these
earlier specific bounds were not derived as part of a general theory.

\section{General envelope theory of Salpeter Hamiltonians}\label{Sec:RET}Let
us now turn to our main topic and consider the spinless-Salpeter Hamiltonian
of Eq.~(\ref{Eq:SH}), $$H=\beta\sqrt{m^2+{\bf p}^2}+V(r)\ ,$$and its
eigenvalues $E.$ We shall assume that the potential $V(r)$ is a smooth
transformation $V(r)=g(h(r))$ of another potential $h(r)$ and that $g$ has
definite convexity so that we~obtain bounds to the energy eigenvalues $E$. We
suppose that the ``basis'' potential $h(r)$ generates~a ``tangential''
Salpeter problem$${\cal H}=\beta\sqrt{m^2+{\bf p}^2}+vh(r)\ ,$$for which the
eigenvalues $e(v),$ or bounds to them, are known. We shall follow here
as~closely as possible the development in Refs.~\cite{Hall83,Hall84,Hall93}
for the corresponding Schr\"odinger problem. We express our results in the
form of two theorems.\begin{enumerate}\item The approximations we shall use
from Sec.~\ref{Sec:CHOP}, regarded as functions of the coupling~$v,$ and also
the (unknown) energy functions $e(v)$ of the ``tangential'' Salpeter
problem~are all {\it concave\/}: $e''(v)<0.$ The latter result represents the
principal claim of Theorem~1.\item In Theorem~2 we begin by using an envelope
representation for the potential $V(r)$~and then demonstrate that all the
spectral formulas that follow possess a certain
structure.\end{enumerate}Finally, as an application, we specialize to the
case of pure power-law ``basis'' potentials~$h(r)$ and, more particularly, to
the Coulomb potential and the harmonic-oscillator potential for which, at
this time, we have spectral information (cf.\ the discussions in
Secs.~\ref{Subsec:CP} and \ref{Subsec:HOP}).

\subsection{Convexity of the energy function}We begin by proving\\[1ex]{\bf
Theorem 1\ } {\it The function $e(v)$ is concave, that is,
$e''(v)<0.$}\\[1ex]{\bf Proof\ } Suppose the exact eigenvalue and
(normalized) eigenvector for the problem posed~by ${\cal
H}=\beta\sqrt{m^2+{\bf p}^2}+vh(r)$ are $e(v)$ and $\psi(v,r).$ By
differentiating $(\psi,{\cal H}\psi)$ with respect~to~$v$ we find
$e'(v)=(\psi,h\psi).$ If we now apply $\psi(v,r)$ as a trial vector to
estimate the energy~of~the operator $\beta\sqrt{m^2+{\bf p}^2}+uh(r),$ in
which $v$ has been replaced by $u,$ we obtain an upper bound to $e(u)$ which
may be written in the form $e(u)\leq e(v)+(u-v)e'(v).$ This inequality
tells~us that the function $e(u)$ lies beneath its tangents; that is to say,
$e(u)$ is {\it concave\/}.\hfill$\Box$

\subsection{The principal envelope formula}With the help of Theorem 1 we are
able to prove\\[1ex]{\bf Theorem 2 (principal envelope formula)\ } {\it
Suppose that the operator $\beta\sqrt{m^2+{\bf p}^2}+vh(r)$ has the exact
lowest eigenvalue $e(v),$ and suppose that the operator $\beta\sqrt{m^2+{\bf
p}^2}+g(h(r))$ has the exact lowest eigenvalue $E.$
Then\begin{equation}E\approx{\cal E}\equiv\min_{v>0}[e(v)-ve'(v)+g(e'(v))]\
.\label{Eq:PEF}\end{equation}If $g$ is concave (that is, $g''<0$), then
$E\le{\cal E};$ if $g$ is convex (that is, $g''>0$), then $E\ge{\cal
E}.$}\\[1ex]

\newpage\noindent{\bf Proof\ } All the tangential potentials we shall employ
have the form $V^{({\rm t})}(r)=a(t)h(r)+b(t),$ where, as in the
Schr\"odinger case, the coefficients $a(t)$ and $b(t)$ are given
by$$a(t)=\frac{V'(t)}{h'(t)}=g'(h(t))\ ,\quad
b(t)=V(t)-a(t)h(t)=g(h(t))-g'(h(t))h(t)\ ,$$and $r=t$ is the point of contact
of the potential $V(r)$ and its tangent $V^{({\rm t})}(r)$. If, for~the~sake
of definiteness, we assume that $V=g(h)$ with $g$ concave (i.e., $g''<0$), we
obtain a family~of upper bounds given by $$E\le\varepsilon(t)=e(a(t))+b(t)\
.$$The best of these is given by optimizing over $t$:$$E\le\varepsilon(\hat
t)=e(a(\hat t))+b(\hat t)\ ,$$where $\hat t,$ the value of $t$ which
optimizes these bounds, is to be determined as the
solution~of$$e'(g'(h(\hat{t})))=h(\hat{t})\ .$$In the spirit of the Legendre
transformation \cite{Gelfand} we now consider another problem which~has the
same solution; this second problem is the one that provides us with our
basic~eigenvalue formula. We consider$${\cal
E}=\min_{v>0}[e(v)-ve'(v)+g(e'(v))]\ ,$$which is well defined since $e(v)$ is
concave. The solution has the critical point $\hat v=g'(e'(\hat v)).$ If we
now apply the correspondence $h(\hat t)=e'(\hat{v}),$ it follows that the
critical point $\hat v$ becomes $\hat v=g'(h(\hat t))$ and the
tangential-potential coefficients $a$ and $b$ become\begin{equation}a(\hat
t)=g'(e'(v))=v\ ,\quad b(\hat t)=g(e'(v))-ve'(v)\ ,\quad v=\hat v\
.\label{Eq:TPC}\end{equation}Meanwhile the original critical (energy) value
is given by$$\varepsilon(\hat t)=e(a(\hat t))+b(\hat t)=e(v)-ve'(v)+g(e'(v))\
,\quad v=\hat v\ .$$\hfill$\Box$\\From the proof of Theorem 2 it follows
immediately that, if the {\it exact\/} energy function $e(v)$ corresponding
to the basis potential $h$ is not available, then, for $g(h)$ concave,
concave~{\it upper\/} approximations $e_{\rm u}(v)>e(v)$ or, for $g(h)$
convex, concave {\it lower\/} approximations $e_{\rm l}(v)<e(v)$ may be used
instead of the exact energy function $e(v)$ in the principal envelope
formula~(\ref{Eq:PEF}). Then all the lower tangents will lie even lower and
all the upper tangents will lie even~higher. If $g$ is convex, we obtain a
lower bound; if $g$ is concave, we obtain an upper bound; because~of the
concavity of $e(v),$ {\it this\/} extremum is a minimum in {\it both\/}
cases. If we wish to use numerical solutions to the ``basis'' problem
(generated by $h(r)$), or if a completely new energy-bound expression becomes
available, the principal envelope formula (\ref{Eq:PEF}) is what would be
used~first.

Interestingly, in the formula (\ref{Eq:PEF}) the tangential-potential
apparatus is no longer evident; only the correct convexity is required. As in
the Schr\"odinger case \cite{Hall83}, once we have the basic result, the
reformulation in terms of ``kinetic potentials'' is often useful: the kinetic
potential $\bar h(s)$ corresponding to some basis potential $h(r)$ is given
by the Legendre transformation~\cite{Gelfand}$$\bar h(s)=e'(v)\ ,\quad
s=e(v)-ve'(v)\ .$$Meanwhile the envelope approximation has the
kinetic-potential expression $\bar{V}(s)\approx g(\bar h(s)).$

For both the Coulomb lower bounds (\ref{Eq:Herbst-bound}) or (\ref{Eq:NLB})
and the harmonic-oscillator upper bounds (\ref{Eq:SEVF-HO}) which we have at
present, we may express our general results in a special common form which
will now be derived.

\subsection{The Coulomb lower bound}We consider first the Coulomb lower bound
in which we assume that the potential $V(r)$~is~a convex transformation
$V(r)=g(h(r))$ of the Coulomb potential $h(r)=-1/r.$ According~to
Sec.~\ref{Subsec:CP}, in this case all the ``lower'' $e_{\rm l}(v)$ have been
arranged---with the parameters~$\beta$~and $m$ returned---in the form$$e_{\rm
l}(v)=\beta m\sqrt{1-\left(\frac{\sigma v}{\beta}\right)^2}\ .$$From this it
follows by elementary algebra that if we define a new optimization
variable~$r$~by $e_{\rm l}'(v)=h(r)=-1/r,$ we have$$e_{\rm l}(v)-ve_{\rm
l}'(v)=\beta\sqrt{m^2+\frac{P^2}{r^2}}\ ,\quad P\equiv\frac{1}{\sigma}\
.$$Consequently, the lower bound on the energy eigenvalues $E$ of the
spinless Salpeter equation becomes\begin{equation}
E\ge\min_{r>0}\left[\beta\sqrt{m^2+\frac{P^2}{r^2}}+V(r)\right],\quad v<\beta
v_P\ .\label{Eq:SSE-LB}\end{equation}Here the boundary value $v_P$ of the
Coulomb coupling $v$ is given, when simply determined by the requirement of
boundedness from below of the operator (\ref{Eq:SH}), by the critical
coupling~$v_{\rm c}$,$$v_P=v_{\rm c}=\frac{2}{\pi}\ ,$$and, when arising from
the region of validity of our Coulomb-like family of lower
bounds~(\ref{Eq:NLB}), via $P=1/\sigma,$ by\begin{equation}
v_P=P\sqrt{1-P^2}<\frac{1}{2}\ .\label{Eq:CCUBvP}\end{equation}$\{P,v_P\}$
pairs may be easily generated from the upper bound on the coupling $v$
in~Eq.~(\ref{Eq:CCUBvP}). The meaning of the Coulomb-coupling constraint is
$a(\hat t)<\beta v_P,$ where $a$ is the coefficient~in the tangential Coulomb
potential given by (\ref{Eq:TPC}).

\subsection{The harmonic-oscillator upper bounds}Next, let us turn to the
harmonic-oscillator upper bounds. Our main assumption is here that
$V(r)=g(r^2),$ with $g''<0.$ In this case the only difficulty is that the
basis problem~$h(r)=r^2$ is equivalent to a Schr\"odinger problem whose
solution ${\cal E}_{n\ell}(v)$ is not known exactly. Following the discussion
after the proof of Theorem 2, let us call the upper bound provided by
Eq.~(\ref{Eq:SEVF-HO}) $e_{\rm u}(v)$ and let us introduce the shorthand
notation $P_{n\ell}(2)=2n+\ell-\frac{1}{2}=P.$ Then we have~the following
parametric equations for $e_{\rm u}(v)$:$$e_{\rm
u}(v)=v\frac{P^2}{r^2}+\beta\sqrt{m^2+r^2}\ ,\quad v=\frac{\beta
r^4}{2P^2\sqrt{m^2+r^2}}\ ,\quad e_{\rm u}'(v)=\frac{P^2}{r^2}\ .$$By
substituting these expressions into the fundamental envelope formula
(\ref{Eq:PEF}) we obtain~the following upper bound on all the eigenvalues of
the spinless-Salpeter problem with potential $V(r)=g(r^2)$ and $g''<0$:
\begin{equation}
E_{n\ell}\le\min_{r>0}\left[\beta\sqrt{m^2+\frac{P^2}{r^2}}+V(r)\right],\quad
P=P_{n\ell}(2)=2n+\ell-\frac{1}{2}\ .\label{Eq:SSE-UB}\end{equation}

\section{The Coulomb-plus-linear (or ``funnel'') potential}\label{Sec:FP}In
order to illustrate the above results by a physically motivated example, let
us apply~these considerations to the Coulomb-plus-linear or (in view of its
shape) ``funnel'' potential$$V(r)=-\frac{c_1}{r}+c_2r\ ,\quad c_1\ge 0\
,\quad c_2\ge 0\ .$$(This potential provides a reasonable overall description
of the strong interactions of quarks in hadrons. For the phenomenological
description of hadrons in terms of both nonrelativistic and semirelativistic
potential models, see, e.g., Refs.~\cite{Lucha91:BSQ,Lucha92:QAQBS}.) By
choosing as basis potential $h(r)=-1/r,$ we may write $V(r)=g(h(r))$ with
$$g(h)=c_1h-\frac{c_2}{h}\ ,$$which is clearly a convex function of $h<0$:
$g''>0.$ Thus the convexity condition is~satisfied. However, we are not free
to choose the coupling constants $c_1$ and $c_2$ as large as we please.~It is
immediately obvious that, for a particular $\{P, v_P\}$ pair, we must in any
case have~$c_1<\beta v_P.$ For the full problem the coefficient $c_2$ of the
linear term will also be involved. The coupling~$v$ we are concerned about is
given by (\ref{Eq:TPC}). We have $$v=g'(e'(v))=\frac{\beta
P^2}{r\sqrt{m^2+\left(\displaystyle\frac{P}{r}\right)^2}}
=c_1+\frac{c_2}{h^2}=c_1+c_2r^2\ .$$From this we obtain, for given values of
the parameters $m$ and $\beta$ and for a given $\{P, v_P\}$~pair, as a
sufficient condition for $v<\beta v_P$ the ``Coulomb coupling constant
constraint'' on the~two coupling strengths $c_1$ and $c_2$ in the funnel
potential:\begin{equation}c_1+\frac{P^2}{m^2}\left(\frac{P^2}{v_P^2}-1\right)
c_2<\beta v_P\ .\label{Eq:CCCC}\end{equation}In the case $\{P=1/\sqrt{2},
v_P=1/2\}$ and $\beta=m=1$ this condition reduces to
$c_1+\frac{1}{2}c_2<\frac{1}{2}.$ For Herbst's lower bound
(\ref{Eq:Herbst-bound}), i.e., $P=v_P=v_{\rm c}=2/\pi,$ this constraint
clearly yields~$c_1<\beta v_P.$ There is no escaping this feature of all
energy bounds involving the Coulomb potential: the constraint derives from
the fundamental observation that the Coulomb coupling $v$ must~not be too
large, so that the (relativistic) kinetic energy is able to counterbalance
the Coulomb potential in order to maintain the Hamiltonian (\ref{Eq:SH}) with
$V(r)=-v/r$ bounded from below.

For example, if we seek the largest allowed value of the parameter $P$ by
solving Eqs.~(\ref{Eq:CCUBvP}) and (\ref{Eq:CCCC}) together, we find that
this largest $P$ is given by
\begin{equation}\frac{c_2\sin^4t}{\cos^2t(\beta\sin t\cos t-c_1)}=m^2\ ,\quad
P\equiv\sin t\ .\label{Eq:P(m)}\end{equation}

\begin{figure}[h]\begin{center}\psfig{figure=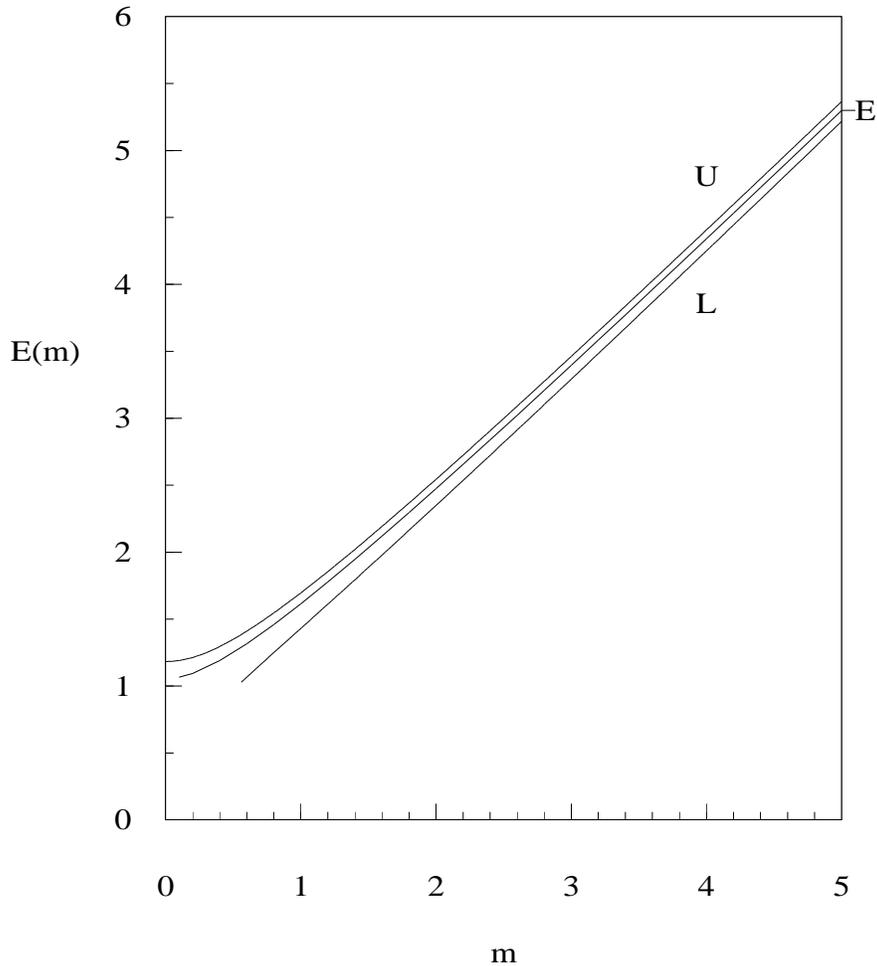,scale=0.79}
\caption{Lower bounds (L), according to (\ref{Eq:SSE-LB}), and upper bounds
(U), according to (\ref{Eq:SSE-UB}),~on the energy eigenvalue $E$ of the
ground state [$(n,\ell)=(1,0)$] of the spinless Salpeter equation with a
Coulomb-plus-linear potential $V(r)=-c_1/r+c_2r,$ for $\beta=1,$ $c_1=0.1,$
and $c_2=0.25.$ The lower bound is given by the general result
(\ref{Eq:SSE-LB}) with the ``best'' $P(m)$ provided~by~(\ref{Eq:P(m)}). In
order to satisfy the Coulomb coupling constraint (\ref{Eq:P(m)}), the mass
$m$ must fulfil $m>\sqrt{5}/4.$ For comparison, a (very accurate)
Rayleigh--Ritz variational upper bound $E$ is
depicted~too.}\label{Fig:Bounds}\end{center}\end{figure}

For the Coulomb-plus-linear potential $V(r)=-c_1/r+c_2r$ under consideration,
Fig.~\ref{Fig:Bounds} shows the lower and upper bounds on the lowest energy
eigenvalue $E$ of the spinless Salpeter equation, given by the envelopes of
the lower and upper families of tangential energy curves (\ref{Eq:SSE-LB})
and (\ref{Eq:SSE-UB}). In the case of the lower bound (\ref{Eq:SSE-LB}), we
have used for each $m$ the best possible $P(m)$ provided by (\ref{Eq:P(m)}).
As $m\to 0,$ the ``basis'' Coulomb problem $H=\beta\sqrt{m^2+{\bf
p}^2}-v/r$~has energy $e(m)\to 0;$ thus the Coulomb lower bound for a
non-Coulomb problem becomes very weak for small values of $m.$ Of course,
Eq.~(\ref{Eq:SSE-UB}) provides us with rigorous upper bounds for {\it
every\/} energy level.

In order to get an idea of the location of the exact energy eigenvalues $E$,
Fig.~\ref{Fig:Bounds} also shows the ground-state energy curve $E(m)$
obtained by the Rayleigh--Ritz variational technique~\cite{Reed78} with the
Laguerre basis states for the trial space defined in Ref.~\cite{Lucha97}.
Strictly speaking, this energy curve represents only an upper bound to the
precise eigenvalue $E$. However,~from~the findings of Ref.~\cite{Lucha97} the
deviations of these Laguerre bounds from the exact eigenvalues~may be
estimated, for the superposition of 25 basis functions used here, to be of
the order of~1\,\%.

\section{Summary and conclusion}In this analysis we have studied the discrete
spectrum of semirelativistic ``spinless-Salpeter'' Hamiltonians $H,$ defined
in Eq.~(\ref{Eq:SH}), by an approach which is based principally on convexity.
We have at our disposal very definite information concerning, on the one
hand, the bottom of the spectrum of $H$ for the Coulomb potential,
$h(r)=-1/r,$ and, on the other hand, the entire spectrum of $H$ for the
harmonic-oscillator potential, $h(r)=r^2.$ The class of potentials that are
at the same time a convex transformation of $-1/r$ and a concave
transformation~of $r^2$ includes, for example, arbitrary linear combinations
of Coulomb, logarithmic, linear,~and harmonic-oscillator terms. The envelope
technique applied here takes advantage of the fact that all ``tangent lines''
to the interaction potential $V(r)=g(h(r))$ in $H$ are potentials~of~the form
$ah(r)+b,$ and that, by convexity and the comparison theorem recalled in
Subsec.~\ref{Subsec:SCT}, the energy eigenvalues corresponding to these
``tangent'' potentials provide rigorous bounds to the unknown exact
eigenvalues $E$ of $H.$ If $e(v)$ denotes the energy function---or a suitable
bound to it---corresponding to the problem posed by a ``basis'' potential
$vh(r),$ where $v$~is~a positive coupling parameter, the envelopes of upper
and lower families of energy curves~may be found with the help of the
``principal envelope formula''$$E\approx\min_{v>0}[e(v)-ve'(v)+g(e'(v))]\
.$$Here, a sign of approximate equality is used to indicate that, for a
definite convexity~of~$g(h),$ the envelope theory yields lower bounds for
convex $g(h)$ and~upper bounds for concave $g(h).$ With the above principal
envelope formula at hand, all new spectral pairs $\{h(r),e(v)\}$ which may
become available at some future time can immediately be used to enrich our
collection of energy bounds. If the basis potential~$h(r)$ is a pure power,
these bounds can be written~as$$E_{n\ell}\approx\min_{r>0}
\left[\beta\sqrt{m^2+\frac{P_{n\ell}^2}{r^2}}+V(r)\right],$$where the numbers
$P_{n\ell}$ are obtained from the corresponding underlying basis
problems.~The power of this technique is illustrated, in Sec.~\ref{Sec:FP},
by our application to the funnel potential, $V(r)=-c_1/r+c_2r.$ For this
problem, we have employed both the semirelativistic Coulomb and
harmonic-oscillator problems to calculate, respectively, lower and upper
bounds on the energy eigenvalues of the spinless Salpeter equation.

We expect that such results would provide bounds on the energy eigenvalues
for general theoretical discussions, or be used as guides for more tightly
focussed analytic or numerical studies of the spectra of semirelativistic
``spinless-Salpeter'' Hamiltonians.

\section*{Acknowledgement}Partial financial support of this work under Grant
No.~GP3438 from the Natural Sciences and Engineering Research Council of
Canada, and the hospitality of the Erwin Schr\"odinger International
Institute for Mathematical Physics in Vienna is gratefully acknowledged by
one of us (R.~L.~H.).

\newpage\end{document}